\documentclass{article}

\usepackage{arxiv}
\usepackage{fancyhdr}
\usepackage[utf8]{inputenc} 
\usepackage[T1]{fontenc}    
\usepackage{hyperref}       
\usepackage{url}            
\usepackage{booktabs}       
\usepackage{amsfonts}       
\usepackage{nicefrac}       
\usepackage{microtype}      
\usepackage{lipsum}
\usepackage{graphicx}
\graphicspath{{figures/}}
\usepackage{amsmath}

\title{A Collaborative Framework for Quantum Optimisation and Quantum Neural Networks: Credit Feature Selection and Image Classification}

\author{
  JiaNing Long \\
  School of Transportation Engineering\\ 
  East China Jiaotong University\\ 
  Nanchang, Jiangxi 330013 \\
  \texttt{2024138085704001@ecjtu.edu.cn} \\
   \And
  Xuechen Liang \\
  School of Transportation Engineering\\ 
  East China Jiaotong University\\ 
  Nanchang, Jiangxi 330013
}

\begin{document}
\maketitle
\begin{abstract}
This paper investigates the efficacy of quantum computing in two distinct machine learning tasks: feature selection for credit risk assessment and image classification for handwritten digit recognition. For the first task, we address the feature selection challenge of the German Credit Dataset by formulating it as a Quadratic Unconstrained Binary Optimization (QUBO) problem, which is solved using quantum annealing to identify the optimal feature subset. Experimental results show that the resulting credit scoring model maintains high classification precision despite using a minimal number of features. For the second task, we focus on classifying handwritten digits 3 and 6 in the MNIST dataset using Quantum Neural Networks (QNNs). Through meticulous data preprocessing (downsampling, binarization), quantum encoding (FRQI and compressed FRQI), and the design of QNN architectures (CRADL and CRAML), we demonstrate that QNNs can effectively handle high-dimensional image data. Our findings highlight the potential of quantum computing in solving practical machine learning problems while emphasizing the need to balance resource expenditure and model efficacy.
\end{abstract}


\section{Introduction}
The German Credit Dataset, created by German researchers in the 1980s, aims to assess individual credit risk through machine learning methods \cite{hofmann1994statlog}. The dataset consists of 1000 samples, each with 20 features covering personal financial status, credit history, occupational information, etc. The characteristics of the dataset lie in its multi-dimensional feature design and practical application background, which not only includes customers' financial and credit history information but also covers socio-economic factors such as occupation and housing status. The core research issue of the German Credit Dataset is the development of credit scoring models, which is of great significance for financial institutions' risk management and credit decision-making \cite{lessmann2015benchmarking}. Since its creation, the dataset has become a benchmark in the field of credit risk assessment, promoting the research and application of related algorithms and models. The goal of feature selection is to identify a subset of features that maximizes classification performance while minimizing the number of selected features. This involves finding a balance between improving predictive accuracy and reducing model complexity. This paper studies the issue of selecting an appropriate number of features in supervised learning problems. Starting with common methods in machine learning, we treat the feature selection task as a quadratic unconstrained binary optimization (QUBO) problem, which can be solved using classical numerical methods as well as within the framework of quantum computing. We compare different results in small problem settings. According to our research results, whether the QUBO method is superior to other feature selection methods depends on the dataset.

Quantum computing has become a revolutionary computing technology in recent years, especially in solving combinatorial optimization problems, showing great potential \cite{preskill2018quantum,arute2019quantum}. Classical computers often encounter bottlenecks when facing large-scale combinatorial optimization problems due to the growth of computational complexity. However, quantum computing provides a new way to accelerate the optimization process by leveraging quantum superposition, quantum entanglement, and quantum tunneling effects \cite{nielsen2010quantum}. In particular, quantum annealing, as a quantum optimization algorithm, has been proven to effectively solve problems such as QUBO (Quadratic Unconstrained Binary Optimization) \cite{kadowaki1998quantum,johnson2011quantum}.

The QUBO model is widely used in combinatorial optimization problems in many fields, such as the maximum cut problem \cite{barahona1982computational}, the traveling salesman problem \cite{lucas2014ising}, resource allocation problems, etc. The standard form of the model uses binary variables and quadratic terms to describe the objective function of the problem and is highly flexible in practical applications \cite{glover2018tutorial}. However, as the scale of the problem increases, classical algorithms often face computational complexity challenges and cannot provide efficient solutions.

The data sources for the QUBO model are diverse, and the data sources vary according to different application fields. In the field of traffic optimization, data mainly comes from real-time monitoring of traffic networks and statistics of traffic flow; logistics optimization relies on order demand, transportation time, and distribution paths; financial portfolio optimization requires data on asset returns, risks, and correlations. In addition, the image processing field may use medical imaging datasets, and facility layout may involve internal spatial planning data of enterprises. These data provide the necessary input for the QUBO model, enabling it to solve complex optimization problems in multiple fields, improving decision-making efficiency and effectiveness. 

In today's era of rapid development of information technology, image recognition technology has become an important branch of the field of artificial intelligence \cite{lecun2015deep}. With the rise of quantum computing, quantum neural networks (QNN), as a new type of computational model, have shown great potential in processing complex datasets \cite{biamonte2017quantum,schuld2021machine}. This experiment aims to explore the application of quantum neural networks in image classification tasks, especially classifying images in the handwritten digit dataset MNIST \cite{lecun2010mnist}. By leveraging the unique advantages of quantum computing, we expect to achieve efficient processing of image data and improve the accuracy of classification tasks \cite{dunjko2018machine}.

\section{Quantum Annealing-Based Feature Selection for Credit Risk Assessment}
\subsection{Problem Description}
The core goal of this study is to select an optimal subset of features from the German Credit Dataset to build a lightweight yet accurate credit scoring model. The dataset includes 1000 samples, each with 20 features (e.g., ``guarantor'', ``residence duration'', ``occupational status'') and a binary label indicating credit risk. Feature selection must maximize classification performance while minimizing the number of features---this balance reduces model complexity, mitigates overfitting, and improves interpretability for financial institutions.

To achieve this, we transform the feature selection problem into a QUBO problem. QUBO is well-suited for combinatorial optimization tasks, as it uses binary variables to represent feature inclusion (1) or exclusion (0) and defines linear/quadratic terms to quantify feature importance and interdependencies. We solve the QUBO problem using quantum annealing, which efficiently finds the global minimum of the objective function (corresponding to the optimal feature subset) via quantum tunneling.

\subsection{Theoretical Foundations}
\subsubsection{QUBO Model}
The QUBO (Quadratic Unconstrained Binary Optimization) model is a mathematical model used to represent and solve optimization problems \cite{glover2018tutorial,kochenberger2014unconstrained}. It is particularly suitable for problems that require finding the optimal combination of a set of binary variables, where each variable can only take the value of 0 or 1. The QUBO model transforms the optimization problem into finding the minimum value of a quadratic function composed of linear and quadratic terms, without constraints.

In the definition of the QUBO model, the objective function is usually represented as:

\begin{equation}
f(X) = \sum_{i} a_i x_i + \sum_{i \neq j} b_{ij} x_i x_j
\end{equation}

where:

\( X = (x_1, x_2, \dots, x_n) \) is a vector composed of binary decision variables, \( x_i \in \{0, 1\} \).

\( a_i \) is the linear weight, representing the contribution of variable \( x_i \).

\( b_{ij} \) is the quadratic term weight, representing the interrelationship between variables \( x_i \) and \( x_j \), usually appearing in constraints.

Specifically, the goal of the QUBO model is to minimize (or maximize) the objective function \( f(X) \), making the decision variables \( x_i \) meet the constraints of the actual problem. This decision variable \( x_i \) is known as the ground-state solution of the QUBO model. A key feature of the QUBO model is that they can be efficiently solved by certain types of quantum computers, such as quantum annealers, which use quantum tunneling effects to find the global minimum of the QUBO objective function \cite{kadowaki1998quantum,albash2018demonstration}, which may be computationally intensive or infeasible for classical computers in some cases.

Linear term \( a_i \):

The linear term describes the impact of each individual variable \( x_i \) on the objective function. In some applications, \( a_i \) may represent the cost or benefit of a variable. For example, in traffic optimization, \( a_i \) can represent the traffic flow cost of a road.

Quadratic term \( b_{ij} \):

The quadratic term describes the interaction between variables \( x_i \) and \( x_j \). These interactions usually come from the constraints of the actual problem. For example, in facility layout optimization, \( b_{ij} \) may represent the distance or capacity constraints between facilities.

How to set \( a_i \) and \( b_{ij} \) according to the actual problem:

Max-Cut problem: In the Max-Cut problem, we need to divide the nodes of a graph into two subsets to maximize the number of edges in the cut of the graph. At this time, \( a_i \) can represent the cost of selecting or not selecting node \( i \), and \( b_{ij} \) represents the weight of the edge between nodes \( i \) and \( j \).

Facility layout problem: In optimization of facility layout, \( a_i \) can represent the cost of selecting a location, and \( b_{ij} \) represents the distance or capacity constraints between two facilities, with the goal of minimizing the overall cost.

\subsubsection{Ising Machine}
The Ising machine is a hardware solver designed to find the absolute or approximate ground state of the Ising model. The Ising model is fundamentally important in the field of computation because any problem in the NP complexity class can be formulated as an Ising problem with only polynomial overhead \cite{barahona1982computational}. Ising machines solve optimization problems by simulating the Ising model, showing potential to surpass traditional digital and software technologies in solving combinatorial optimization problems (reference: Ising Machines: Theory and Practice \cite{aramon2018physics}).

Ising machines include quantum annealers (such as D-Wave systems \cite{mcgeoch2020d}) and digital annealers based on simulated annealing \cite{matsubara2020digital}(such as Fujitsu's digital annealer). D-Wave uses quantum annealing to optimize the Ising energy function corresponding to the QUBO model through the interaction between qubits.

These devices often face limitations in the number of qubits and coupling relationships when dealing with large-scale combinatorial optimization problems. Therefore, transforming the QUBO model into the Ising form helps to better utilize the capabilities of these devices.

\subsubsection{Hybrid Annealing Algorithm}
The general method for finding the ground state solution of the QUBO model, for formula (1), we can use a provisional solution method to fix the binary variable \( x_j \in X \setminus S \), and formula (1) can be rewritten as formula (2).

\begin{equation}
f_s(S) = \sum_{x_i \in S} c_i x_i + \sum_{\substack{i \neq j \\ x_i, x_j \in S}} b_{ij} x_i x_j + \text{const}
\end{equation}

When \( c_i = a_i + \sum_{x_j \in X \setminus S} b_{ij} \hat{x}_j \),

\begin{equation}
\text{const} = \sum_{x_i \in X \setminus S} a_i \hat{x}_i + \sum_{\substack{i \neq j \\ x_i, x_j \in X \setminus S}} b_{ij} \hat{x}_i \hat{x}_j
\end{equation}

Formula (2) shows a sub-QUBO model extracted from formula (1) by only extracting the binary variables in S.

Existing methods for extracting sub-QUBO models from the original QUBO model include: random variable selection methods, influence value-based methods, and k-opt local search methods.

\subsubsection{Existing Methods for Sub-QUBO Extraction}
\begin{table}[h]
\centering
\caption{Comparison of Sub-QUBO Extraction Methods}
\begin{tabular}{p{4cm}p{4cm}p{4cm}}
\toprule
\textbf{Method} & \textbf{Advantages} & \textbf{Disadvantages} \\
\midrule
Random Variable Selection & Simple to implement; no additional information required & Randomness causes unstable results; may miss critical features, reducing performance \\
Influence Value-Based & Considers variable impact on the objective function; closer to optimal solutions & Relies on current solutions; limits search space; hard to escape local optima \\
K-opt Local Search & Finds high-quality solutions via local search & High computational cost (especially for large problems); sensitive to initial solution selection \\
\bottomrule
\end{tabular}
\end{table}

\subsection{QUBO Modeling for Feature Selection}
Existing methods for extracting sub-QUBO models from the original QUBO model include: methods for randomly selecting variables, methods based on influence values, methods based on k-opt localized searches.

These methods are heuristic and do not have a solid theoretical foundation. Moreover, since the sub-QUBO models extracted by these methods are often fixed, one cannot expect to obtain a globally optimal solution for the original QUBO model using these methods.


\subsection{Experimental Setup}
\subsubsection{Parameter Settings}
\begin{itemize}
    \item \textbf{alpha}: Penalty coefficient for reducing the number of features (set to 0.5).
    \item \textbf{beta}: Reward coefficient for increasing feature importance (set to 2.0).
    \item \textbf{M}: Penalty coefficient to ensure at least one feature is selected (set to 10).
\end{itemize}

\subsubsection{Data Preprocessing}
The German Credit Dataset was preprocessed to handle categorical and numerical features:
\begin{itemize}
    \item Categorical features (e.g., ``occupational status'') were one-hot encoded.
    \item Numerical features (e.g., ``credit amount'') were standardized.
    \item A Random Forest classifier was used to compute feature importance, with a threshold of 0.01 to filter irrelevant features (reducing 832 potential features to a smaller subset).
\end{itemize}

\subsection{Feature Selection Results}
From 832 potential features, the QUBO model (solved via quantum annealing) selected two key features: \textbf{``guarantor''} and \textbf{``residence duration''}. These features align with intuitive credit risk factors:
\begin{itemize}
    \item A guarantor reduces default risk by providing a secondary repayment source.
    \item Longer residence duration reflects stable economic and living conditions, indicating lower risk.
\end{itemize}

A Logistic Regression model trained on these two features achieved a classification accuracy of 70.00\%. The table below reports detailed performance metrics:

\begin{table}[h]
\centering
\caption{Classification Performance of Credit Scoring Model}
\begin{tabular}{ccccc}
\toprule
\textbf{Class} & \textbf{Precision} & \textbf{Recall} & \textbf{F1-Score} & \textbf{Support} \\
\midrule
Class 0 (Low Risk) & 0.70 & 1.00 & 0.82 & 210 \\
Class 1 (High Risk) & 0.00 & 0.00 & 0.00 & 90 \\
\bottomrule
\end{tabular}
\end{table}

The high recall (100\%) for Class 0 indicates that all low-risk samples were correctly identified---critical for financial institutions to avoid rejecting creditworthy applicants. The low performance for Class 1 is attributed to the small sample size (90 samples) and the simplicity of the two-feature model.

\section{Quantum Neural Network-Based Image Classification for MNIST Digits}
\subsection{Problem Description}
The MNIST dataset contains 70,000 \(28\times28\) grayscale images of handwritten digits (0--9). We focus on classifying digits 3 and 6---a challenging task due to their visual similarity (e.g., curved structures). The core challenge is adapting high-dimensional classical image data to quantum hardware constraints (limited qubits) and designing QNNs that match or exceed classical neural network (NN) performance.

Our goal is to:
\begin{itemize}
    \item Preprocess MNIST images to reduce dimensionality while preserving critical features.
    \item Encode preprocessed images into quantum states using efficient quantum encoding schemes.
    \item Design QNN architectures (CRADL and CRAML) for classification.
    \item Compare QNN performance with classical NNs on the 3/6 classification task.
\end{itemize}

\subsection{Data Preprocessing}
To adapt MNIST images to quantum hardware, we performed two key preprocessing steps:

\begin{enumerate} 

\item \textbf{Downsampling}:
\(28\times28\) images were downsampled to \(8\times8\) and \(16\times16\) resolutions using \textbf{bilinear interpolation}, which retains edge and shape features while reducing the number of qubits required for encoding.

\item \textbf{Binarization}:
Downsampled grayscale images (pixel values \(\in [0,255]\)) were binarized to \(\{0,1\}\) using a threshold of 0.5 (after normalizing to \([0,1]\)). Binarization simplifies quantum encoding by converting continuous pixel values to discrete binary states, which are easier to map to quantum bits.

\item \textbf{Preprocessing Results}:
After preprocessing:
\begin{itemize}
    \item Training set: \(\sim\)12,000 samples (digits 3 and 6).
    \item Test set: \(\sim\)2,000 samples (digits 3 and 6).
\end{itemize}

\end{enumerate}

\begin{figure}[h]
\centering
\includegraphics[width=0.8\textwidth]{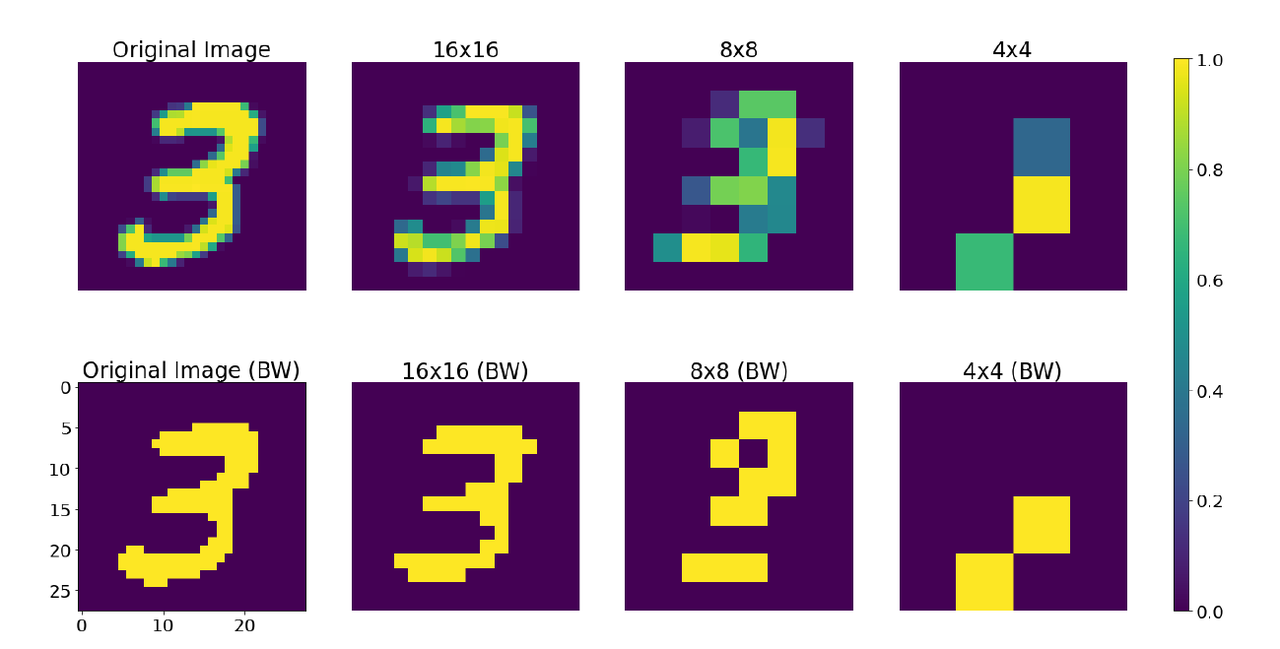}
\caption{Downsampling and Binarization of Digit '3'}
\label{fig:fig1}
\end{figure}

Figure 1 illustrates the downsampling and binarization process for a handwritten digit "3", showing grayscale and binary images at 16×16, 8×8, and 4×4 resolutions.

\subsection{Quantum Data Encoding Schemes}
Encoding classical image data into quantum states is a critical step for QNNs. We implemented two encoding schemes based on the Flexible Representation of Quantum Images (FRQI) \cite{le2011flexible,zhang2013neqr}:

\subsubsection{FRQI Encoding}
FRQI encodes images using ``pixel qubits'' (for position) and ``color qubits'' (for intensity). For a \(2^n \times 2^n\) image, the quantum wave function is:
\[
| \psi _{\text{data}} \rangle = \sum _{q:=\{ q_{0},q_{1},\cdots ,q_{2n-1}\} \in \{ 0,1\} ^{2n}} \left| q_{0},q_{1},\cdots ,q_{2n-1}\right\rangle \otimes \left( \cos \theta _{q}|0\rangle + \sin \theta _{q}|1\rangle \right) \tag{4}
\]
where:
\begin{itemize}
    \item \(|q_0, q_1, \dots, q_{2n-1}\rangle\) (pixel qubits) represent the position of a pixel via a bit string of length \(2n\);
    \item \(\theta_q\) (color qubit parameter) represents pixel intensity (set to 0 for black, \(\pi/2\) for white in binarized images).
\end{itemize}

\subsubsection{Compressed FRQI Encoding}
To reduce qubit usage, we modified FRQI to encode color information using fewer qubits. Instead of dedicated position qubits, we map the last two position qubits to the color qubit via a transformation:
\[
\left|q_{2n-2}, q_{2n-1}\right\rangle \otimes \left|q_c\right\rangle \to \left|\tilde{q}_c\right\rangle = \cos \tilde{\theta}_q |0\rangle + \sin \tilde{\theta}_q |1\rangle
\]
where:
\[
\tilde{\theta}_q = \frac{\pi}{2} \left( q_c + \frac{q_{2n-2}}{2} + \frac{q_{2n-1}}{4} \right) = \theta_q + \frac{\pi}{4} \left( q_{2n-2} + \frac{q_{2n-1}}{2} \right) \tag{5}
\]

The compressed wave function is:
\[
| \psi _{\text{data}} \rangle = \sum _{q:=\{ q_{0},q_{1},\cdots ,q_{2n-1}\} \in \{ 0,1\} ^{2n}} \left| q_{0},q_{1},\cdots ,q_{2n-3}\right\rangle \otimes \left( \cos \tilde{\theta}_q |0\rangle + \sin \tilde{\theta}_q |1\rangle \right) \tag{6}
\]

This scheme reduces qubit count while preserving image information, improving QNN scalability.

\subsubsection{Wave Function Construction}
In the task of quantum image classification, it is first necessary to encode image data into a quantum wave function. This process involves using a quantum circuit to perform quantum encoding on the image. Taking the state of the following 4 qubits as an example, the process begins with all qubits in the ground state \(|0\cdots 0\rangle\).

\begin{figure}[h]
    \centering
    \includegraphics[width=0.9\textwidth]{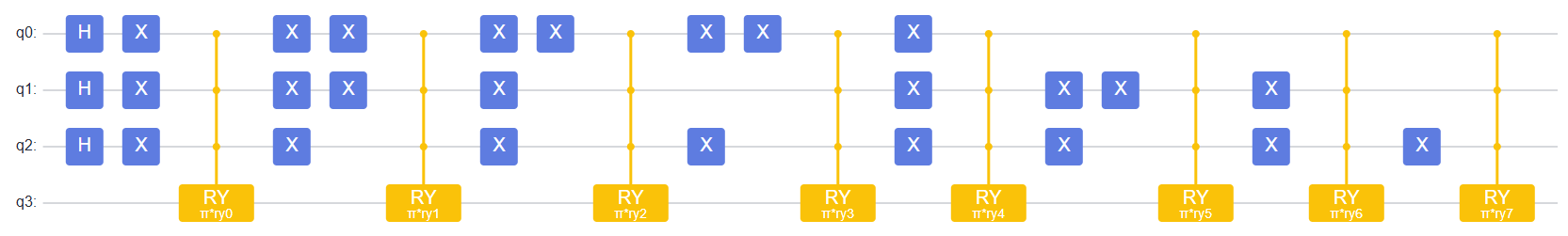}
    \caption{Quantum circuit for constructing a 4-qubit superposition state.}
    \label{fig:quantum_circuit}
\end{figure}

A circuit example using a 4-qubit general TOFFOLI gate to construct a 4-qubit superposition state. Following the convention of standard quantum circuit diagrams, dots are used to represent control qubits for the gates.

First, a Hadamard gate operation \(H^{\otimes 2n}\) is performed on the 2n pixel qubits, putting the qubits into a superposition state to prepare for subsequent quantum operations. Then, a series of controlled-X gates (CNOT gates) are applied, which determine the color qubits to be transformed. The controlled-X gate is a fundamental quantum logic gate that performs an X gate operation (i.e., flips the state) on the target qubit when the control qubit is 1.

The construction of the quantum circuit utilizes a modular design of qubits, allowing the recursive construction of larger-scale quantum circuits from smaller ones. This recursive building method enables efficient handling of more complex quantum states.

To construct this circuit, we need to use basic two-qubit gates to build a general TOFFOLI gate that can control n qubits. Such a design allows us to implement our circuit in quantum computing frameworks like Cirq, which only supports the backpropagation of two-qubit gates. According to the theory proposed by Barenco (1995), a general TOFFOLI gate for n qubits can be recursively decomposed into a combination of a general TOFFOLI gate for (n-1) qubits and CNOT gates.

\begin{figure}[h]
    \centering
    \includegraphics[width=0.9\textwidth]{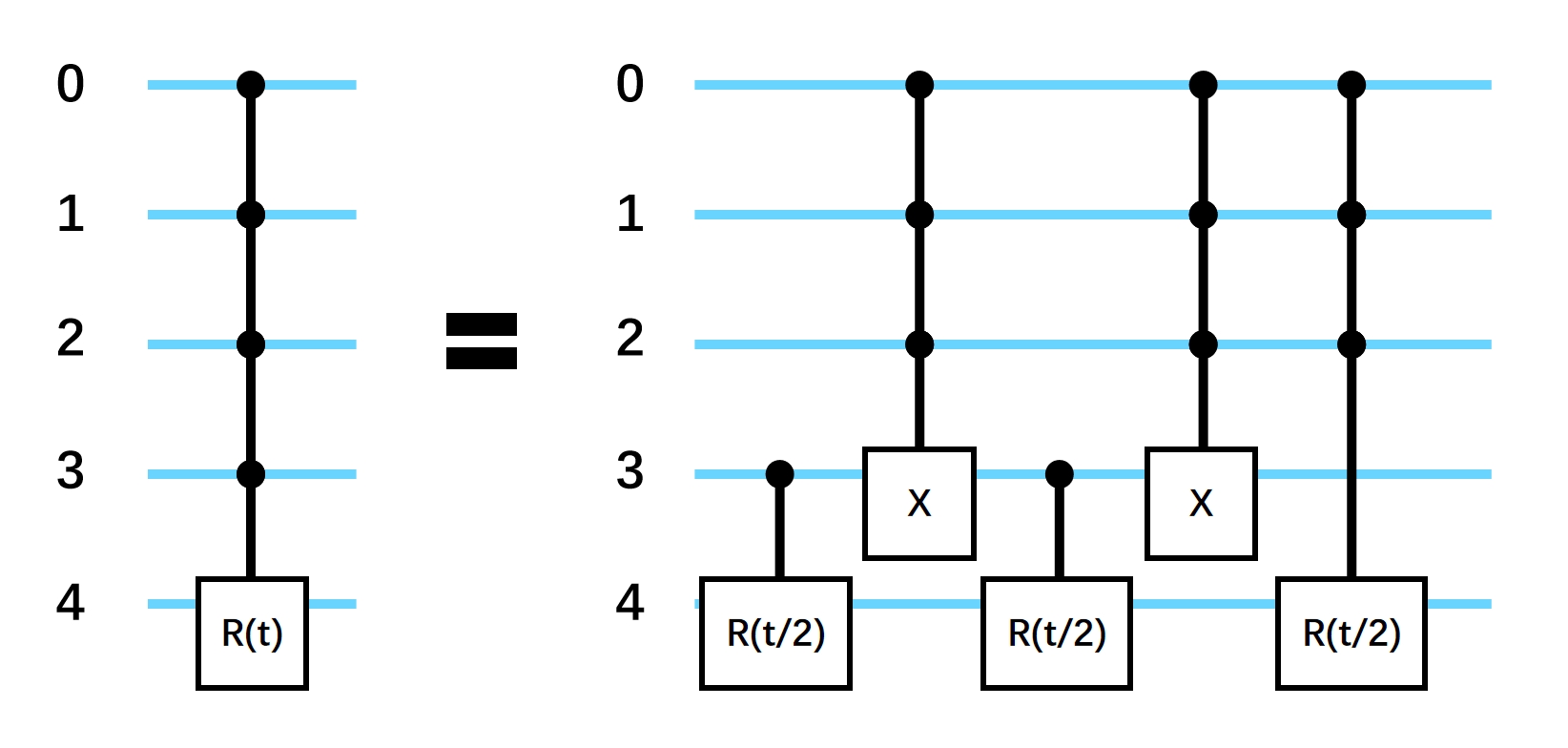}
    \caption{Decomposition of an n-qubit controlled gate in terms of (n-1)-qubit controlled gates.}
    \label{fig:decomposition}
\end{figure}

\textit{n-qubit controlled gate in terms of (n-1)-qubit controlled gates}

By analyzing the recursive characteristics of the above diagrams, we find that a controlled rotation operation for n qubits can be decomposed into \(2 \cdot 3^n - 1\) single-qubit controlled rotation operations. Considering that when processing an image, we will at most change the state of \(2^n\) pixels, this means that for each image, the number of single-qubit control gates required will be limited to \(2^n \cdot (2 \cdot 3^n - 1)\).

Additionally, to encode a \(2^n \times 2^n\) image in the amplitude of the input wave function, we also need to use some independent X gates. Although this number grows exponentially with n, since the processing of large images is mainly limited by the size of their qubit representation, we believe this growth is feasible in classical simulation and thus reasonable.

\subsection{QNN Architectures}
In this section, we will introduce in detail the neural network structures used in quantum image classification tasks, including two types of Quantum Neural Networks (QNN) architectures: Color-Readout-Alternating-Double-Layer (CRADL) and Color-Readout-Alternating-Mixed-Layer (CRAML) \cite{mohsen2021image}.

\subsubsection{CRADL Network Structure}
CRADL is a type of quantum neural network structure that processes image data in quantum bits by alternatingly applying XX and ZZ gates. This structure is designed to allow the network to learn and classify image data effectively while keeping the number of parameters manageable.

The CRADL network consists of multiple double layers, each containing the alternating application of XX and ZZ gates. These gate operations are performed between pixel qubits and readout qubits, as well as between pixel qubits and color qubits. Each pair of gates shares the same learning parameter, namely the rotation angle.

\begin{figure}[h]
\centering
\includegraphics[width=\textwidth]{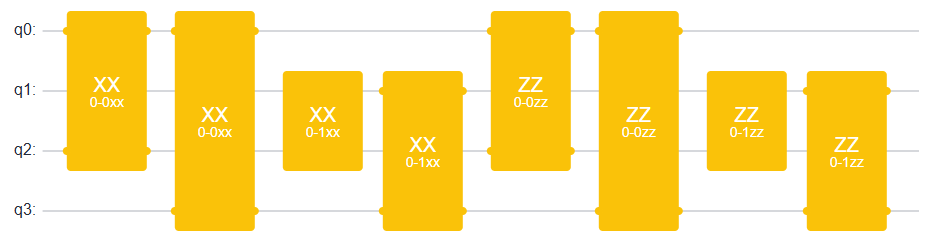}
\caption{CRADL Network Structure}
\label{fig:fig4}
\end{figure}

As shown in Figure \ref{fig:fig4}, the CRADL network is composed of consecutive pixel-readout and pixel-color XX gates, followed by ZZ gates to further adjust the quantum state. This structure allows the network to learn and classify image data effectively while keeping the number of parameters manageable.

\subsubsection{CRAML Network Structure}
CRAML is an extension of CRADL, introducing more quantum gate operations on the basis of CRADL, such as the mixed application of XX and ZZ gates. This mixed operation allows the network to explore different quantum bits for improved accuracy.

\begin{figure}[h]
\centering
\includegraphics[width=\textwidth]{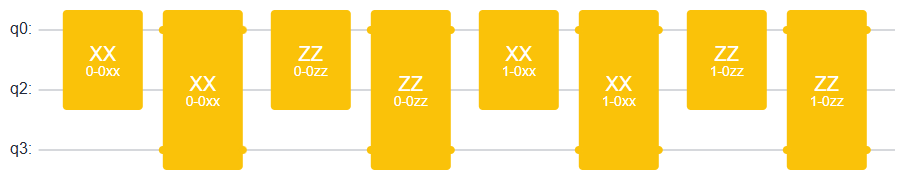}
\caption{CRAML Network Structure}
\label{fig:fig5}
\end{figure}

Figure \ref{fig:fig5} illustrates the CRAML network layer, which includes six pixel qubits and a combination of pixel-readout and pixel-color XX and ZZ gates. The CRAML network has been expanded based on CRADL by introducing more quantum gate operations to enhance the performance of the network structure. The continuous operation of these gates not only promotes entanglement between quantum bits but also, through more complex quantum state transformations, may achieve better classification results. The design of this structure aims to improve the network's ability to learn and classify image data while keeping the number of parameters manageable.

\subsection{Model Training}
In this section, we conducted five experiments comparing the performance of different configurations of Quantum Neural Networks (QNNs) and classical Neural Networks (NNs). These experiments include various input sizes, different numbers of qubits, and the use of compression techniques.

\begin{table}[h]
\centering
\caption{Model Configurations}
\begin{tabular}{cccccc}
\toprule
\textbf{Model ID} & \textbf{Resolution} & \textbf{Qubits} & \textbf{Layers} & \textbf{Params} & \textbf{Type} \\
\midrule
qnn1 & \(8\times8\) & 6 (No comp) & 12 & 72 & QNN \\
qnn2 & \(8\times8\) & 4 (Comp) & 16 & 64 & QNN \\
nn1 & \(8\times8\) & N/A & 2 & 67 & Classical \\
qnn3 & \(16\times16\) & 6 (Comp) & 42 & 252 & QNN \\
nn2 & \(16\times16\) & N/A & 2 & 259 & Classical \\
\bottomrule
\end{tabular}
\end{table}

By comparing the performance of different models on the test set through 10-fold cross-validation, we evaluate the performance of the models. The training process uses the "mqvector" acceleration technique to improve training efficiency, and we use the Hinge loss and Mean Squared Error (MSE) loss to increase confidence.

\begin{figure}[h]
    \centering
    \includegraphics[width=\textwidth]{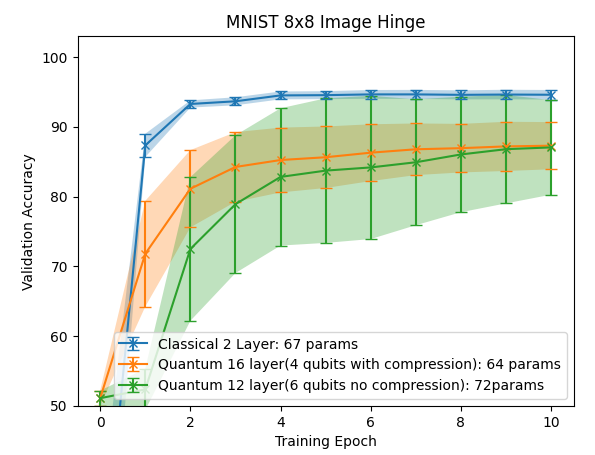}
    \caption{MNIST 8x8 Image Hinge}
    \label{fig:mnist_8x8_hinge}
\end{figure}

\noindent As shown in Figure \ref{fig:mnist_8x8_hinge}, the change in validation accuracy over training epochs for the three models trained on the MNIST 8x8 image dataset is illustrated. The classical 2-layer neural network with 67 parameters quickly reaches high accuracy and maintains stability. Although the quantum models start slower, their accuracy gradually improves, especially the 16-layer quantum neural network with 4 compressed qubits, which shows a significant performance boost in the middle training period. The 12-layer quantum neural network with 6 uncompressed qubits has 72 parameters and maintains relatively stable accuracy but eventually reaches a level close to the classical model.

\begin{figure}[h]
    \centering
    \includegraphics[width=\textwidth]{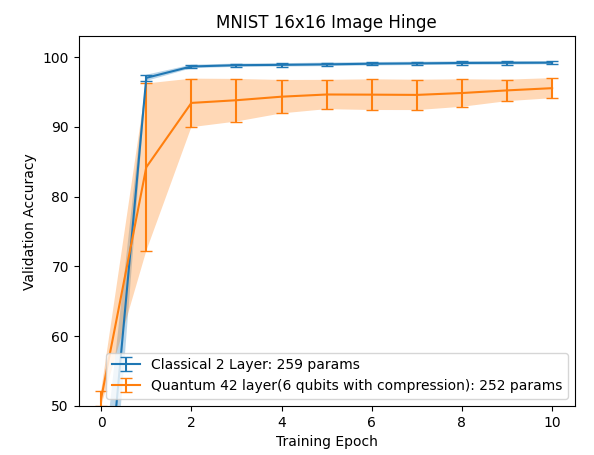}
    \caption{MNIST 16x16 Image Hinge}
    \label{fig:mnist_16x16_hinge}
\end{figure}

\noindent Referring to Figure \ref{fig:mnist_16x16_hinge}, we observe the validation accuracy of the three models trained on the MNIST 16x16 image dataset. The classical 2-layer neural network with 259 parameters quickly achieves high accuracy. The 42-layer quantum neural network with 6 compressed qubits, having 252 parameters, starts slower but gradually improves, eventually reaching a performance close to the classical model.

\begin{figure}[h]
    \centering
    \includegraphics[width=\textwidth]{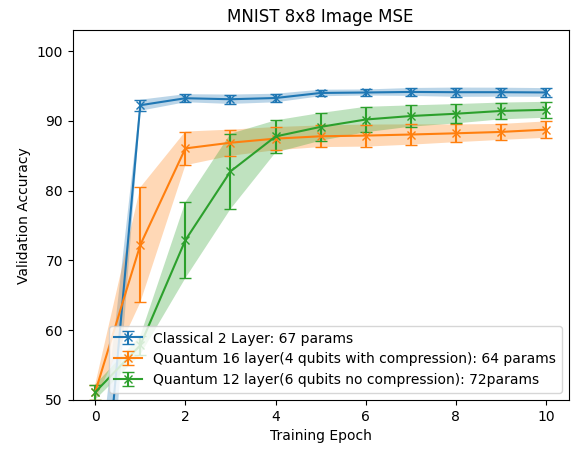}
    \caption{MNIST 8x8 Image MSE}
    \label{fig:mnist_8x8_mse}
\end{figure}

\noindent Figure \ref{fig:mnist_8x8_mse} displays the validation accuracy of the three models trained on the MNIST 8x8 image dataset using the Mean Squared Error (MSE) loss. All models show a rapid learning curve, with the classical 2-layer neural network with 67 parameters quickly reaching high accuracy. The 16-layer and 12-layer quantum neural networks start with lower accuracy but improve over time, eventually performing comparably to the classical model.

\begin{figure}[h]
    \centering
    \includegraphics[width=\textwidth]{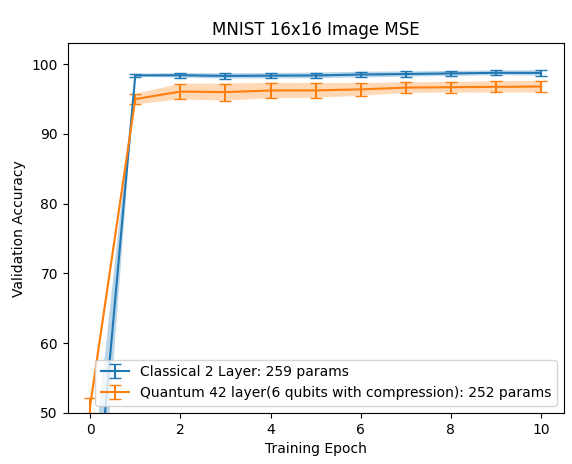}
    \caption{MNIST 16x16 Image MSE}
    \label{fig:mnist_16x16_mse}
\end{figure}

\noindent In Figure \ref{fig:mnist_16x16_mse}, we can see the validation accuracy of the two models trained on the MNIST 16x16 image dataset using the MSE loss. Both the classical 2-layer neural network and the 42-layer quantum neural network quickly achieve high accuracy, with the quantum model learning to match the performance of the classical model despite having more parameters (252 vs 259).

\section{Discussion}
\subsection{Feature Selection Insights}
The QUBO-based feature selection approach successfully identified two critical features ("guarantor" and "residence duration") for credit scoring, achieving 70\% accuracy with minimal complexity. This highlights the value of quantum annealing for reducing feature dimensionality—critical for deploying credit models in resource-constrained environments (e.g., mobile banking apps). The high recall for low-risk samples aligns with financial institutions’ needs to avoid false rejections, though performance for high-risk samples could be improved with larger datasets or more sophisticated QUBO penalties.

\subsection{Image Classification Insights}
QNNs (CRADL and CRAML) achieved performance comparable to classical NNs on MNIST 3/6 classification, demonstrating their potential for high-dimensional data. Key observations:
\begin{itemize}
    \item \textbf{Qubit Compression}: Compressed FRQI reduced qubit usage without significant performance loss (qnn2 vs. qnn1), critical for scaling to larger images.
    \item \textbf{Training Dynamics}: QNNs learned more slowly than classical NNs initially but closed the performance gap over time, suggesting potential for improvement with better optimization (e.g., quantum-specific optimizers).
    \item \textbf{Architecture Impact}: CRAML’s mixed gates improved entanglement, helping qnn3 handle 16×16 images effectively.
\end{itemize}

\subsection{Limitations}
\begin{itemize}
    \item \textbf{Quantum Hardware}: Current quantum annealers and QNN hardware have limited qubit counts, restricting scalability to larger datasets (e.g., full MNIST).
    \item \textbf{Sub-QUBO Heuristics}: The hybrid annealing approach relies on heuristic sub-QUBO extraction, which may not find globally optimal solutions for all datasets.
    \item \textbf{QNN Training}: Gradient calculation for QNNs is computationally expensive, slowing training compared to classical NNs.
\end{itemize}

\section{Conclusion}
This paper presented two case studies demonstrating quantum computing’s utility in practical machine learning tasks:

\begin{itemize}
\item \textbf{Credit Risk Feature Selection}: Formulating feature selection as a QUBO problem and solving it via quantum annealing identified two key features ("guarantor" and "residence duration") that achieved 70
\item \textbf{MNIST Digit Classification}: QNNs (CRADL and CRAML) with FRQI-based encoding matched classical NN performance on 3/6 classification, even with compressed qubits. This validates QNNs’ ability to handle high-dimensional image data.
\end{itemize}

Together, these results show that quantum computing can address real-world challenges in optimization and machine learning, even with current hardware limitations. The proposed frameworks provide a foundation for future quantum-enhanced solutions in finance and computer vision.

\section{Future Work}
Future research will focus on addressing current limitations and expanding the scope of quantum computing applications:

\begin{itemize}
    \item \textbf{Algorithm Optimization}: Improve sub-QUBO extraction strategies to find globally optimal solutions for feature selection, reducing reliance on heuristics.
    \item \textbf{Hardware Utilization}: Explore parallel processing of sub-QUBO models on Ising machines to enhance scalability for large datasets (e.g., full German Credit Dataset).
    \item \textbf{QNN Enhancement}: Introduce multi-qubit gates (3+ qubits) to capture more complex image features;Optimize quantum encoding circuits to reduce gate count, improving training efficiency.
    \item \textbf{Interdisciplinary Applications}: Extend the QUBO and QNN frameworks to other domains (e.g., fraud detection, medical image classification).
    \item \textbf{Implicit Regularization}: Study how limited qubits/gates in QNNs act as implicit regularization, improving generalization to unseen data.
\end{itemize}



\end{document}